\documentclass[twocolumn,aps,prb]{revtex4}
\usepackage{graphicx}
\usepackage{amssymb}
\usepackage{amsmath}
\usepackage{bm}

\def\Journal #1,#2,#3,#4#5#6#7{#1 {\bf #2}, #3 (#4#5#6#7)}

\def\Vec#1{\bm{#1}}
\def\vare{\varepsilon}

\def\p{\prime}
\def\o{\omega}
\def\ve{\varepsilon}

\def\d{\delta}
\def\rm{\mathrm}
\def\H{\mathcal H}
\def\t{\theta}

\begin{document}

\title{Orbital magnetic susceptibility of finite-sized graphene}
\author{Yuya Ominato and Mikito Koshino}
\affiliation{Department of Physics, Tohoku University, Sendai 980-8578, Japan}
\date{\today}

\begin{abstract}
We study the orbital magnetism of graphene ribbon
in the effective-mass approximation,
to figure out the finite-size effect 
on the singular susceptibility known in the bulk limit.
We find that the susceptibility at $T=0$
oscillates between diamagnetism and paramagnetism
as a function of $\vare_F$, in accordance with the subband structure
formed by quantum confinement.
In increasing $T$, the oscillation rapidly disappears 
once the thermal broadening energy exceeds the subband spacing, 
and the susceptibility $\chi(\vare_F)$ approaches the bulk limit
i.e., a thermally broadened  diamagnetic peak centered at $\vare_F=0$.
The electric current supporting the diamagnetism is found to
flow near the edge with a depth 
 $\sim \hbar v /(2\pi k_B T)$, with $v$ being the band velocity,
while at $T=0$ the current distribution spreads entirely in the sample
reflecting the absence of the characteristic wavelength in graphene.
The result is applied to estimate the three-dimensional
random-stacked multilayer graphene, where we show that
the external magnetic field is significantly screened 
inside the sample in low temperatures,
in a much stronger manner than in graphite.
\end{abstract}

\maketitle

\section{INTRODUCTION}

Graphite is known as one of the strongest diamagnetic materials
among natural substances.
\cite{McClure_1956a,McClure_1960a,Sharma_et_al_1974a}
This property is due to the large orbital 
diamagnetism related to the small effective mass
in the band structure,
i.e., narrow energy gap between conduction and valence bands.
The diamagnetic effect becomes even greater 
in graphene monolayer
\cite{Novoselov_et_al_2004a,Novoselov_et_al_2005a,Zhang_et_al_2005a} 
which is truly a zero-gap system.
\cite{McClure_1956a,Slonczewski_and_Weiss_1958a,DiVincenzo_and_Mele_1984a,Semenoff_1984a}
The magnetic susceptibility of graphene at zero temperature
contains a singularity expressed as a delta
function in Fermi energy $\varepsilon_F$,
which diverges at Dirac point ($\vare_F=0$) where
the two bands stick, and vanishes otherwise.
\cite{McClure_1956a,Sharapov_et_al_2004a,Fukuyama_2007a,Nakamura_2007a,Koshino_and_Ando_2007b,Ghosal_et_al_2007a,Ando_2007d,Koshino_et_al_2009a,Koshino_and_Ando_2010a}
The orbital diamagnetism 
has been studied for other graphene-related materials, 
such as graphite intercalation compounds,
\cite{Safran_and_DiSalvo_1979a,Safran_1984a,
Blinowski_and_Rigaux_1984a,Saito_and_Kamimura_1986a} 
carbon nanotube,
\cite{Ajiki_and_Ando_1993b,Ajiki_and_Ando_1995c,Yamamoto_et_al_2008a} 
few-layer graphenes,
\cite{Koshino_and_Ando_2007c,Nakamura_and_Hirasawa_2008a,
Castro_Neto_et_al_2009a}
and an organic material having similar gapless spectrum. 
\cite{Kobayashi_et_al_2008a}

In this paper, we investigate the orbital diamagnetism
of a graphene strip with finite width. 
\cite{Fujita_et_al_1996a,Nakada_et_al_1996a,Wakabayashi_et_al_1999a,Ezawa_2006a,Brey_and_Fertig_2006a,Brey_and_Fertig_2006b,Son_et_al_2006a,Son_et_al_2006b,Han_et_al_2007a,Chen_et_al_2007a,Li_et_al_2008a,Kosynkin_et_al_2009a,Jiao_et_al_2009a}
The purpose of this work is two-fold:
(i) To understand how the delta-function
singularity of the bulk limit is relaxed in a realistic 
finite-sized graphene system. 
In the literatures, the orbital susceptibility
of graphene nanoribbons was calculated 
for the Fermi energies near the Dirac point,
\cite{Wakabayashi_et_al_1999a,Liu_et_al_2008a} 
while the behavior off Dirac point and 
the relation to the bulk susceptibility is not well understood.
(ii) To study the diamagnetic current flow on graphene.
In the conventional diamagnetism of metal, we  
usually expect that the current circulates near the surface
with a depth of the order of the Fermi wave length $\lambda_F$.
In graphene, the only characteristic length scale $\lambda_F$ 
intrinsically diverges, and we 
expect a peculiar current distribution
different from the conventional system. 

To address above problems, here we calculate the orbital susceptibility
and the current distribution of graphene ribbon with an arbitrary width, 
in various Fermi energies $\vare_F$ and temperatures $T$,
using the effective mass approximation.
We find that the susceptibility at $T=0$
oscillates between diamagnetic and paramagnetic values
in increasing $\vare_F$, 
in accordance with the detailed subband structure.
In increasing temperature,  the oscillation rapidly disappears 
once the thermal broadening energy exceeds
the subband spacing, 
and the susceptibility approaches bulk limit, i.e.,
a thermally-broadened diamagnetic peak centered at $\vare_F=0$,
independently of the atomic configuration at the edge.
We also apply a similar analysis to the carbon nanotube,
and find a similar oscillation in the susceptibility.

The electric current supporting the diamagnetism 
spreads entirely in the sample at $T=0$,
reflecting the absence of the characteristic wavelength.
In increasing temperature, however, 
the current density tends to localize 
near the boundary with a depth  $\sim \hbar v /(2\pi k_B T)$,
forming an edge current circulation.

The analysis of the spatial distribution of the 
diamagnetic current is useful
in studying a graphene stack
where the diamagnetic current of one layer influences 
the electron motion in other layers.
If we take a randomly-stacked graphene multilayer,
in which the interlayer coupling is expected to be small, 
\cite{Berger_et_al_2006,Hass_et_al_2007a,G.Li_et_al_2009a,Lopes_dos_Santos_et_al_2007a,Latil_et_al_2007a,Shallcross_et_al_2008a,Mele_2010a}
the self-consistent calculation shows 
that the diamagnetism is much stronger than in graphite, and
the external magnetic field is significantly screened 
inside the sample in low temperatures.

The paper is organized as follows.
In Sec.\ \ref{sec_form}, we briefly review 
effective mass description of electrons in
graphene ribbon, and formulate the orbital magnetic susceptibility.
We present the numerical results and detailed discussion 
in Sec.\ \ref{sec_num}, as well as
a similar analysis for the carbon nanotube
in Sec.\ \ref{sec_cnt}.
We argue the diamagnetism of randomly stacked graphene multilayers
in Sec.\ \ref{sec_random}.
The conclusion is given in Sec.\ \ref{sec_concl}.

\begin{figure}[t]
\begin{center}
\leavevmode\includegraphics[width=0.9\hsize]{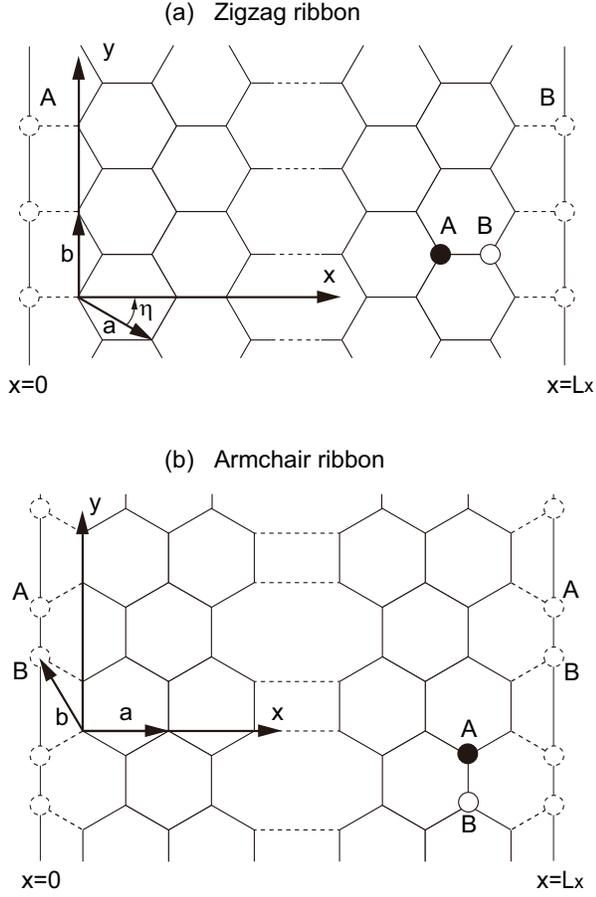}
\end{center}
\caption{Atomic structures of graphene ribbons
with (a) zigzag boundary and (b) armchair boundary,
respectively.
Dashed circles indicate missing sites beyond the boundary.}  
\label{fig_ribbon}
\end{figure}

\section{Formulations}
\label{sec_form}

\subsection{Effective mass approximation}

Graphene is composed of a honeycomb network of carbon atoms, 
where a unit cell contains a pair of sublattices, denoted by $A$ and $B$.
Fig.\ref{fig_ribbon} (a) and (b) show the atomic 
structure of zigzag and armchair graphene ribbons, respectively,
where $\bm{a}$ and $\bm{b}$ are primitive translation vectors 
of infinite graphene. The lattice constant is
given by $a = |\Vec{a}| \approx 0.246$ nm.
For both cases we set $y$-axis along the ribbon, 
and set $x=0$ and $L_x$ to the line of missing sites nearest from the edge.
We define $\eta$ as the angle between $x$ axis and $\bm{a}$,
which is $\pi/6$ for zigzag, and $0$ for armchair boundary.

In a tight-binding model, the wave function of graphene  electron is written as
\begin{align}
\psi(\bm{r})=\sum_{\bm{R}_\rm{A}}\psi_\rm{A}(\bm{R}_\rm{A})\phi(\bm{r}-\bm{R}_\rm{A})+\sum_{\bm{R}_\rm{B}}\psi_\rm{B}(\bm{R}_\rm{B})\phi(\bm{r}-\bm{R}_\rm{B}) ,
\end{align}
where $\Vec{R}_\textrm{A}$ and $\Vec{R}_\textrm{B}$ are the positions of
A-sites and B-sites, respectively,
and $\phi(\bm{r})$ denotes the wave function of the 
$p_z$ orbital of a carbon atom.

For states in the vicinity of the Fermi level $\ve=0$,
the wave amplitudes are written as \cite{Ando_2005a}
\begin{align}
\psi_\rm{A}(\bm{R}_\rm{A})&=e^{i\bm{K}\cdot\bm{R}_\rm{A}}F^{\rm{K}}_{\rm{A}}(\bm{R}_\rm{A})
                                        +e^{i\eta}e^{i\bm{K}^\p\cdot\bm{R}_\rm{A}}F^{\rm{K}^\p}_{\rm{A}}(\bm{R}_\rm{A}),\notag \\
\psi_\rm{B}(\bm{R}_\rm{B})&=-\o e^{i\eta}e^{i\bm{K}\cdot\bm{R}_\rm{B}}F^{\rm{K}}_{\rm{B}}(\bm{R}_\rm{B})
                                        +e^{i\bm{K}^\p\cdot\bm{R}_\rm{B}}F^{\rm{K}^\p}_{\rm{B}}(\bm{R}_\rm{B}) .\label{am}
\end{align}
in terms of the slowly-varying envelope functions $F_{\rm{A}}^{\rm{K}}, F_{\rm{B}}^{\rm{K}}, F_{\rm{A}}^{\rm{K}^{\p}},
\rm{and} {~}F_{\rm{B}}^{\rm{K}^\p}$.
The envelope functions satisfy the Schr\"{o}dinger equation,
\cite{McClure_1956a,Slonczewski_and_Weiss_1958a,DiVincenzo_and_Mele_1984a,Semenoff_1984a,Ando_2005a,Shon_and_Ando_1998a}
\begin{align}
\H_0\bm{F}(\bm{r})=\ve\bm{F}(\bm{r}),\label{Sh} 
\end{align}
with 
\begin{align}
\H_0=\hbar v\begin{pmatrix}
           0 & \hat k_x-i\hat k_y & 0 & 0 \\
           \hat k_x+i\hat k_y & 0 & 0 & 0 \\
           0 & 0 & 0 & \hat k_x+i\hat k_y \\
           0 & 0 & \hat k_x-i\hat k_y & 0 
           \end{pmatrix},
\end{align}

\begin{align}
\bm{F}(\bm{r})=\begin{pmatrix}
                      F^{\rm{K}}_{\rm{A}}(\bm{r}) \\
                      F^{\rm{K}}_{\rm{B}}(\bm{r}) \\
                      F^{\rm{K}^{\p}}_{\rm{A}}(\bm{r}) \\
                      F^{\rm{K}^{\p}}_{\rm{B}}(\bm{r})
                     \end{pmatrix},
\end{align}
where $\hat{\bm{k}}=-i\bm{\nabla}$ and $v$ is the band velocity.

The electronic states of the graphene ribbon can be correctly
described by setting the appropriate boundary condition
to the effective mass Hamiltonian. \cite{Brey_and_Fertig_2006b}
Now the eigenstates are labeled by $k_y$
since the system is translationally symmetric along $y$.
A wave function of $k_y$ and the energy $\vare$ is generally written as
\begin{align}
\bm{F}(\bm{r})=e^{ik_yy}\begin{pmatrix}
                      Ae^{ik_xx}+Be^{-ik_xx} \\
                      s\left(Ae^{i\left(k_xx+\t\right)}-Be^{-i\left(k_xx+\t\right)}\right) \\
                      Ce^{ik_xx}+De^{-ik_xx} \\
                      s\left(Ce^{i\left(k_xx-\t\right)}-De^{-i\left(k_xx-\t\right)}\right)
                     \end{pmatrix},
\label{bs}
\end{align}
where $k_x^2=\ve^2/\hbar^2 v^2-k_y^2,$ 
$e^{i\t}=(k_x+ik_y)/\sqrt{k_x^2+k_y^2}$, $s=\vare/|\vare|$,
and $A,B,C$ and $D$ are numbers to be determined by
satisfying the boundary condition, as we will argue in the following.

\subsection{Zigzag boundary}

In the zigzag ribbon, 
the boundary condition is given by
$\psi_{\rm{A}}(\bm{R}_{\rm{A}})=0$ at $x=0$,
and $\psi_{\rm{B}}(\bm{R}_{\rm{B}})=0$ at $x=L_x$.
By using Eq.\ (\ref{am}), this is translated to
the condition for the envelope function as 
\begin{eqnarray}
F_{\rm{A}}^\rm{K}(0,y)=0,\nonumber \\
F_{\rm{B}}^\rm{K}(L_x,y)=0,\nonumber \\
F_{\rm{A}}^{\rm{K}^{\p}}(0,y)=0,\nonumber \\
F_{\rm{B}}^{\rm{K}^{\p}}(L_x,y)=0, \label{zib}
\end{eqnarray}
which keeps the states at K and those at K$'$ independent.
For an eigenstate for the $\rm{K}$ point,
we apply the first two lines of Eq.(\ref{zib}) 
to Eq.(\ref{bs}), to obtain
\begin{align}
&A+B=0, \notag \\
&s\left(Ae^{i\left(k_xL_x+\t\right)}-Be^{-i\left(k_xL_x+\t\right)}\right)=0.
\end{align}
To have a solution other than $A=B=0$, we require \cite{Brey_and_Fertig_2006b}
\begin{align}
k_y=\frac{k_x}{\tan k_xL_x}.\label{bb}
\end{align}
\begin{figure}[t]
\begin{center}
\leavevmode\includegraphics[width=0.90\hsize]{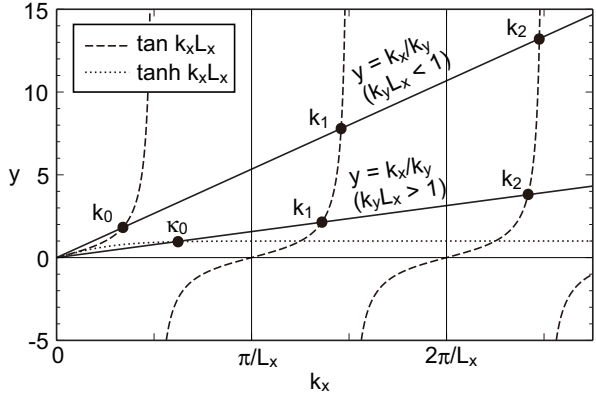}
\end{center}
\caption{Plots to find  
solutions of Eq.(\ref{bb}) and Eq.(\ref{eb}).} \label{xtanxtanhx}
\end{figure}

For given $k_y$, 
we define $k_n (n=0,1,2,\cdots)$ 
as solution of Eq.(\ref{bb}) in $k_x$
satisfying $n\pi<k_nL_x<(n+1)\pi$.
Corresponding eigenstates and the energy are obtained as
\begin{align}
&\bm{F}_{snk_y}(\bm{r})=A_n\frac{e^{{ik_yy}}}{\sqrt{L_xL_y}}\begin{pmatrix}
 i\sin k_nx \\
s(-1)^{n+1}\sin k_n(x-L_x) \\
0 \\ 0
\end{pmatrix}, \notag \\
&A_n=\left(1-\frac{\sin2k_nL_x}{2k_nL_x}\right)^{-1/2}, \notag \\
&\ve_{snk_y}=s\hbar v\sqrt{k_n^2+k_y^2},
\label{es}
\end{align}
for $n=0,1,2,\cdots.$
For the normalization of the wavefunction,
we assumed the periodic boundary condition 
in $y$-direction with a large enough period $L_y$.

Solution $k_n$ is obtained by searching for 
crossing points
of $\tan k_xL_x$ and $k_x/k_y$  as illustrated in Fig.\ \ref{xtanxtanhx}.
When $k_yL_x>1$, the first solution $k_0$ becomes
a pure imaginary number $i\kappa_0$ which satisfies 
\begin{align}
k_y=\frac{\kappa_0}{\tanh \kappa_0 L_x}.\label{eb}
\end{align}
The wavefunction and the energy then becomes
\begin{align}
&\bm{F}_{s0k_y}(\bm{r})=A_1\frac{e^{{ik_yy}}}{\sqrt{L_xL_y}}
\begin{pmatrix}
i\sinh \kappa_0 x \\ 
-s\sinh \kappa_0(x-L_x) \\
0\\ 0 
\end{pmatrix}, \notag \\
&A_0=\left(-1+\frac{\sinh2\kappa_0 L_x}{2\kappa_0 L_x}\right)^{-1/2}, \notag \\
&\ve_{s0k_y}=s\hbar v\sqrt{-\kappa_0^2+k_y^2}.
\label{es2}
\end{align}
This actually describes the edge state 
localized at the boundary $x=0$ and $L_x$
giving a nearly flat energy band.
\cite{Fujita_et_al_1996a,Nakada_et_al_1996a,Wakabayashi_et_al_1999a}

The eigenenergy $\vare_{s n k_y}$ represents 
the $n$-th branch of conduction $(s=+)$
and valence $(s=-)$ bands respectively.
Energy band structure of K as a function of $k_y$ 
is shown as solid curves in Fig.\ref{allmag}(a).
Eigenstates for $\rm{K}^\p$ point are obtained
similarly, where the energy band structure is equivalent to
Fig.\ \ref{allmag}(a) with $k_y$ inverted to $-k_y$.
The flat band of edge states of K and K$'$ are connected 
in a wave number away from K or K$'$. \cite{Fujita_et_al_1996a,Nakada_et_al_1996a,Wakabayashi_et_al_1999a}

\subsection{Armchair boundary}

In the armchair ribbon, 
the boundary condition imposes
both of $\psi_{\rm{A}}(\bm{R}_{\rm{A}})=0$ 
and $\psi_{\rm{B}}(\bm{R}_{\rm{B}})=0$ 
at each of $x=0$ and $x=L_x$.
The corresponding conditions for the envelope functions are written as
\begin{align}
F_{\rm{A}}^\rm{K}(0,y)+F_{\rm{A}}^{\rm{K}^{\p}}(0,y)=0,\notag \\
F_{\rm{B}}^\rm{K}(0,y)-F_{\rm{B}}^{\rm{K}^{\p}}(0,y)=0,\notag \\
F_{\rm{A}}^\rm{K}(L_x,y)+\o^{-2N}F_{\rm{A}}^{\rm{K}^{\p}}(L_x,y)=0,\notag \\
F_{\rm{B}}^\rm{K}(L_x,y)-\o^{-2N}F_{\rm{B}}^{\rm{K}^{\p}}(L_x,y)=0,
\end{align}
where $N = L_x/a$ is the number of honeycomb lattices
between $x=0$ and $L_x$, which can be integer or half-integer
depending on the position of the edge.

Applying above conditions to Eq.(\ref{bs}), we obtain
\begin{align}
\begin{pmatrix}
1 & 1 & 1 & 1 \\
e^{i\t} & -e^{-i\t} & -e^{-i\t} & e^{i\t} \\
e^{i\lambda} & e^{-i\lambda} & \alpha e^{i\lambda} & \alpha e^{-i\lambda} \\
e^{i(\lambda+\t)} & -e^{-i(\lambda+\t)} & -\alpha e^{i(\lambda+\t)} & \alpha e^{-i(\lambda+\t)} 
\end{pmatrix} \begin{pmatrix}
                  A \\
                  B \\
                  C \\
                  D
                  \end{pmatrix}=\begin{pmatrix}
                  0 \\
                  0 \\
                  0 \\
                  0
                  \end{pmatrix},\label{ab}
\end{align}
where $\alpha=\o^{-2N}$ and $\lambda=k_xL_x.$
The determinant of matrix in Eq.(\ref{ab}) should vanish to have a
non-zero solution. This condition is reduced to
\begin{align}
k_x = k_n \equiv 
\frac{\pi}{L_x}
\left(n - \frac{\nu}{3}\right), {~}
n=0,\pm1,\pm2,\cdots,
\label{eq_kn_arm}
\end{align}
$\nu$ is an integer $(0,\pm1)$ defined by
\begin{align}
 2N = 3m + \nu,
\end{align}
with integer $m$.
The eigenstate and energy are obtained as
\begin{align}
&\bm{F}_{snk_y}(\bm{r})=\frac{e^{ik_yy}}{2\sqrt{L_xL_y}}\begin{pmatrix}
                                                                                e^{ik_nx} \\
                                                                                se^{i(k_nx+\t)} \\
                                                                                -e^{-ik_nx} \\
                                                                                se^{-i(k_nx-\t)} 
                                                                              \end{pmatrix}, \notag \\
&\ve_{snk_y}=s\hbar v\sqrt{k_n^2+k_y^2}.
\end{align}

\begin{figure*}
\begin{center}
\leavevmode\includegraphics[width=0.85\hsize]{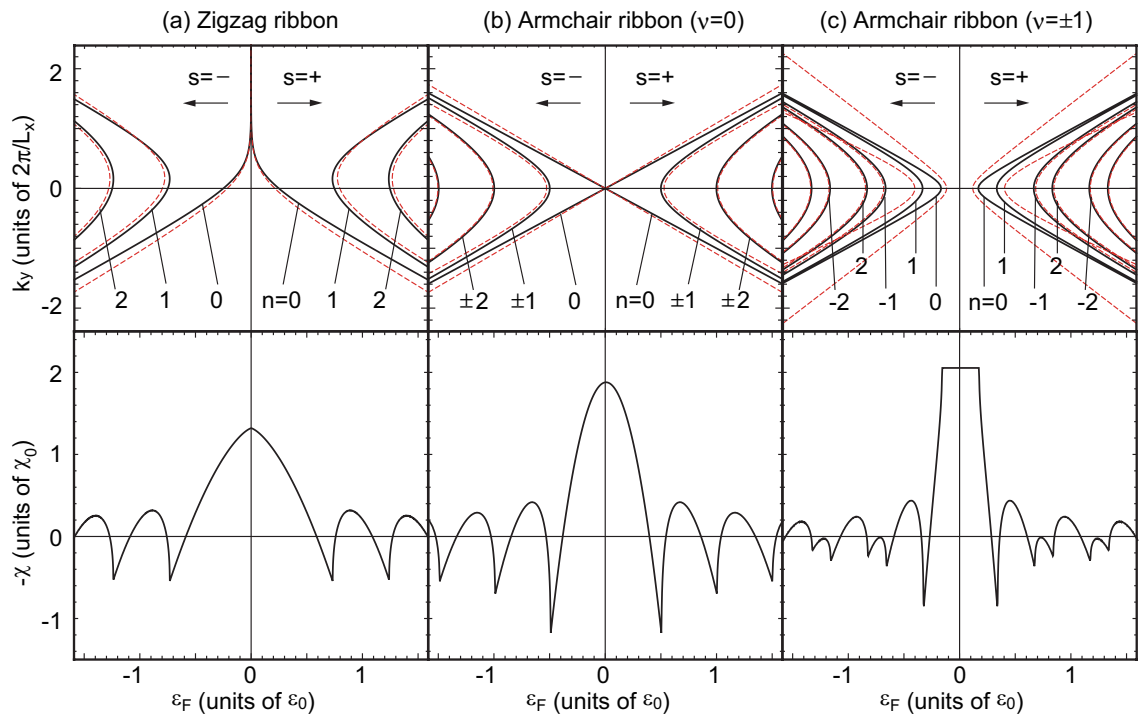}
\end{center}
\caption{Band structure (upper panel)  
and magnetic susceptibility as a function of $\vare_F$ (lower),
of (a) zigzag, (b) metallic armchair $(\nu=0)$ and
(c) semiconducting armchair $(\nu=\pm 1)$ graphene ribbons.
In upper panels, solid (black) and dashed (red) curves 
indicate the band structures 
at zero and a finite magnetic field, respectively.
For the latter, the energy band is calculated 
with the perturbation theory in a magnetic field $B$.
where we take $B=B_0$ for (a), and $B=0.5 B_0$
in (b) and (c) for illustrative purpose,
and $B_0 = (h/e)/L_x^2$.
}
\label{allmag}
\end{figure*}

When $\nu=0$, the energy bands of $n=0$ and $s=\pm$
stick together and thus the system is metallic,
while otherwise a gap opens at zero energy and
the system becomes a semiconductor.
Energy bands for metallic armchair ribbon ($\nu=0$) and 
semiconducting armchair ribbon ($\nu=\pm 1$)
are shown in Fig.\ \ref{allmag}(b) and (c), respectively.
In (c), the labeling $n$ is for the case of $\nu=+1$,
while $n$ becomes $-n$ in $\nu=-1$.

\subsection{Orbital susceptibility}

To calculate the orbital diamagnetism,
we consider a graphene ribbon under 
a uniform magnetic field $B$ perpendicular
to graphene plane.
We take the Landau gauge and set the vector potential as
\begin{align}
\Vec{A}(\Vec{r}) = \left[0, \, B\left(x-\frac{L_x}{2}\right)\right].
\label{eq_A}
\end{align}
The Hamiltonian in presence of the magnetic field
is obtained by replacing $\hat{\bm{k}}$ by
$\hat{\bm{k}}+e\bm{A}/(\hbar c)$, as
\begin{align}
\H&=\H_0+\d\H, \quad \d\H = \frac{e}{c} \hat{v_y} A_y,
\label{eq_H_in_B}
\end{align}
where $c$ is the light velocity, and
\begin{align}
          \hat v_y=\frac{1}{\hbar}\frac{\partial\H}{\partial \hat{k}_y}
=v\begin{pmatrix}
                                                                                      0 & -i & 0 & 0 \\
                                                                                      i & 0 & 0 & 0 \\
                                                                                      0 & 0 & 0 & i \\
                                                                                      0 & 0 & -i & 0 
                                                                                     \end{pmatrix}.
\end{align}

The operator of the electric current density is given by
\begin{align}
\hat j_y(\bm{r})&=-\frac{e}{2}\left\{\hat
 v_y\d(\bm{r}-\bm{r}^{\p})+\d(\bm{r}-\bm{r}^{\p})\hat v_y\right\}.
\label{eq_jy}
\end{align}
In the first order perturbation in $\delta {\cal H}$, 
the expectation value of the  current density is written as
\begin{align}
j_y(\Vec{r}) = 
\sum_{\alpha}
f(\ve_{\alpha})  
\sum_{\beta (\neq \alpha)}
\frac{2\rm{Re}
\left[(\delta\mathcal{H})_{\alpha \beta} (\hat{j}_y(\Vec{r}))_{\beta\alpha}
\right]}
{\ve_{\alpha}-\ve_{\beta}}, \label{cur}
\end{align}
where $\alpha$ and $\beta$ represent
the unperturbed eigenstates of graphene ribbon,
and $f(\vare) = 1/[1+e^{\beta(\vare-\mu)}]$ 
is the Fermi distribution function with the chemical potential 
$\mu$.

In a zigzag ribbon, the current density always vanishes at the edges $x=0$ and 
$L_x$, while it is not generally the case in armchair ribbons. 
This is obvious from at the matrix element of $\hat{j}_y$
between two eigen states $\Vec{F}$ and $\Vec{F}'$,
\begin{align}
 \langle \Vec{F}'|
\hat{j}_y(\Vec{r}) | \Vec{F} \rangle
= i e v 
 \left[F'^K_A(\Vec{r})^* F^K_B(\Vec{r}) - F'^K_B(\Vec{r})^* F^K_A(\Vec{r})
\right].
\label{eq_jy_element}
\end{align}
In the wavefunction of zigzag ribbon,
Eqs.\ (\ref{es}) and (\ref{es2}), the component
$F^K_A$ is zero at $x=0$, and $F^K_B$ is at $L_x$, 
so that Eq.\ (\ref{eq_jy_element}) vanishes at the both edges.

The current density on $xy$-plane is related to
the local magnetic moment $m(\bm{r})$ in $z$-direction
by
\begin{align}
j_x=c\frac{\partial m}{\partial y},{~}j_y=-c\frac{\partial m}{\partial x}.
\end{align}
In the present case, $m(\bm{r})$ depends only on $x$ so that
the total magnetization per area is
\begin{align}
M&=\frac{1}{L_xL_y}\int_0^{L_x} \int_0^{L_y} m(x) \rm{d}x\rm{d}y 
\notag \\
  &=\frac{1}{cL_x}\int_0^{L_x}\left(x-\frac{L_x}{2}\right)j_y(x)\rm{d}x.
\end{align}
The magnetic susceptibility is written as
\begin{align}
\chi&=\lim_{B \to 0}\frac{M}{B} 
     =\frac{2}{cL_xL_y}
\sum_{\alpha}
f(\ve_{\alpha})  
\sum_{\beta (\neq \alpha)}
\frac{\big|(\delta\mathcal{H}/B)_{\alpha \beta}\big|^2}
{\ve_{\beta}-\ve_{\alpha}} . \label{chi}
\end{align}

We calculate Eqs.\ (\ref{cur}) and (\ref{chi}) numerically.
As we have infinite energy bands below zero,
we introduce a cut-off function $g(\vare_\alpha)$
which smoothly vanishes $|\vare_\alpha| > \vare_c$.
In the following, we take $\vare_c = 50 \vare_0$
where 
\begin{equation}
 \ve_0 = \frac{2\pi\hbar v}{L_x}
\end{equation}
is the typical energy scale for the subband structure.
The result is actually converging in the limit of large $\vare_c$.

The susceptibility of the infinite bulk graphene
at zero temperature is given by
\cite{McClure_1956a,Safran_and_DiSalvo_1979a,Koshino_and_Ando_2007b}
\begin{equation}
\chi_\textrm{gr}(\varepsilon_F) = 
-g_vg_s \frac{e^2 v^2}{6\pi c^2} \delta(\varepsilon_F),
\label{eq_chi_zero_T}
\end{equation}
where $g_v=g_s=2$ are the valley and spin degeneracies, respectively.
At finite temperature, it becomes
\begin{align}
\chi_\textrm{gr}(\mu; T) 
&= \int_{-\infty}^\infty d\vare
\left(-\frac{\partial f(\vare)}{\partial \vare}\right)
\chi_\textrm{gr}(\varepsilon) 
\notag\\
&= -g_vg_s \frac{e^2 v^2}{24\pi c^2} \frac{1}{k_BT\cosh[\mu/(2k_B T)]}.
\label{eq_chi_in_T}
\end{align}
In the graphene ribbon,
the characteristic unit of the susceptibility can be chosen as
\begin{equation}
 \chi_0=g_v g_s \frac{e^2v^2}{6\pi c^2}\frac{1}{\vare_0}.
\end{equation}

\begin{figure}
\begin{center}
\leavevmode\includegraphics[width=0.95\hsize]{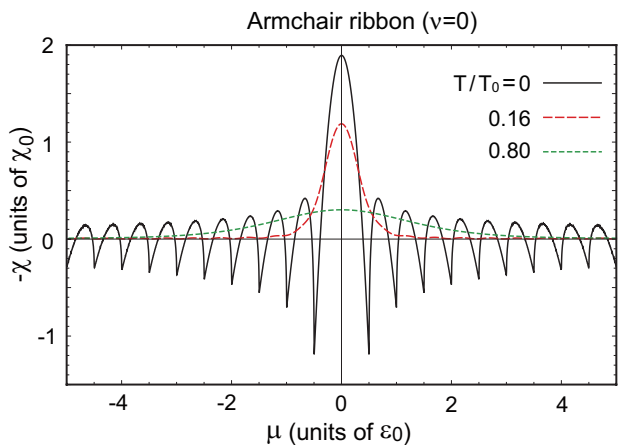} 
\end{center}
\caption{Magnetic susceptibility 
against chemical potential $\mu$
for metallic armchair ribbon $(\nu=0)$ 
at several different temperatures.} 
\label{metalmagboke}
\end{figure}

\section{NUMERICAL RESULTS}
\label{sec_num}

\subsection{Magnetic susceptibility}

Lower panels of Fig.\ref{allmag} show
the orbital susceptibility $\chi(\vare_F)$
of (a) zigzag, (b) metallic armchair  $(\nu=0)$ and (c)
semiconducting armchair  $(\nu=\pm 1)$ ribbons at zero temperature,
where the upward direction represents
the negative (i.e., diamagnetic) susceptibility.
The figures are to be
compared with the band structures in upper panels. 
In every case, the magnitude of $\chi$ becomes the maximum 
at $\vare_F = 0$, and oscillates as a function of $\ve_F$
in accordance with the subband structure.
In the positive energy region, for example,
the curve sharply rises
when a subband starts to be occupied by electrons,
while it tends to decrease otherwise.
In large $|\vare_F|$,
the amplitude of the oscillation slowly attenuates 
approximately in proportional to $1/\sqrt{|\vare_F|}$.

The oscillating feature can be understood in terms of the
band energy shift in an infinitesimal magnetic field.
In the upper figures of Fig.\ \ref{allmag},
we plot as broken curves the energy band in some small $B$ calculated by 
the second order perturbation.
Here the amplitude $B$ is set to some finite value
for illustrative purpose.
Generally the system is diamagnetic
when the total energy shift caused by $B$ is positive,
and paramagnetic when negative.
In the metallic armchair ribbon (b), for example,
we see that a pair of the first subbands $(n=0, s=\pm)$
shift towards zero energy, 
due to the level repulsions from excited subbands nearby.
All other bands $(n>0)$ move in the opposite direction away from zero energy,
while the absolute shifts are much smaller than that of $n=0$.

When $\vare_F$ is zero,
the energy gain of the first valence band $(s=-, n=0)$
exceeds the energy loss of all other valence bands
$(s=-, |n|\geq 1)$,
resulting in the total diamagnetism.
When $\vare_F$ is shifted to positive side,
the diamagnetism decreases
because the first conduction band $(s=+, n=0)$
has a negative shift and gives paramagnetism.
When the second conduction band $(s=+, |n|=1)$
starts to be filled,
the susceptibility suddenly jumps to diamagnetic direction,
because the shift is positive there
and also the density of states diverges at the band bottom.
The oscillation of other types, (a) and (c),
can be explained in a similar manner.


In Fig.\ \ref{metalmagboke},
the susceptibility at several different temperatures 
is plotted as a function of the chemical potential $\mu$.
We here choose the metallic armchair ribbon $(\nu=0)$
while the qualitative property is the same in other cases.
We define the characteristic temperature scale $T_0$ as
\begin{equation}
k_B T_0 = \vare_0,
\end{equation}
at which the thermal broadening energy is of the order of
the subband interval energy.
We see that the oscillation  rapidly disappears
once $T$ becomes of the order of $T_0$,
leaving only single diamagnetic peak at $\vare_F=0$.
When $T >\sim T_0$, the curve becomes almost identical
with the bulk susceptibility, Eq.\ (\ref{eq_chi_in_T}).
We also confirmed that the integration of $\chi$ over $\vare_F$
is identical with the bulk value $-g_vg_s e^2v^2/(6\pi c^2)$ 
within the numerical accuracy,
for all the types of ribbons considered here.
This fact suggests that the finite size effect 
disappears $T >\sim T_0$, regardless of the edge configuration.

\begin{figure}
\begin{center}
\leavevmode\includegraphics[width=0.95\hsize]{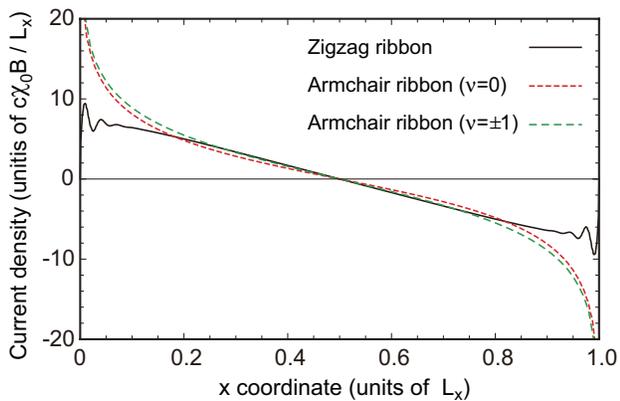}
\end{center}
\caption{
Diamagnetic current density $j_y(x)$
of different types of graphene ribbons
with $\vare_F=0$ at $T=0$.
} \label{allcur}
\end{figure}

\begin{figure}
\begin{center}
\leavevmode\includegraphics[width=0.9\hsize]{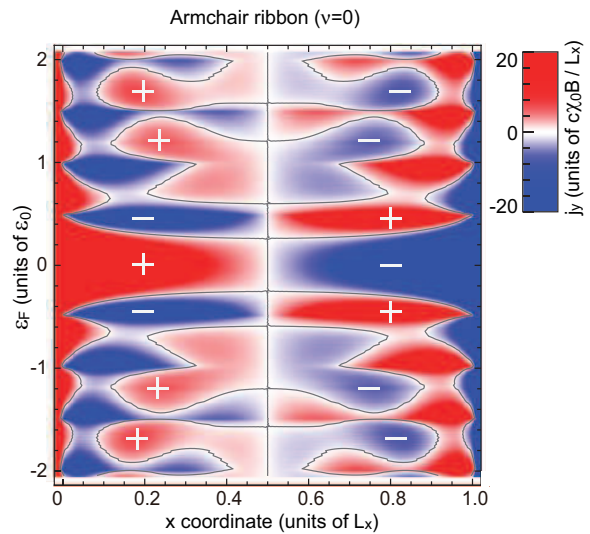}
\end{center}
\caption{
Two-dimensional density plot $j_y(x; \vare_F)$
of the diamagnetic current density of metallic armchair ribbon
$(\nu=0)$, as a function of position $x$ (horizontal axis)
and  Fermi energy (vertical).}
\label{cur3d}
\end{figure}

\begin{figure}
\begin{center}
\leavevmode\includegraphics[width=0.95\hsize]{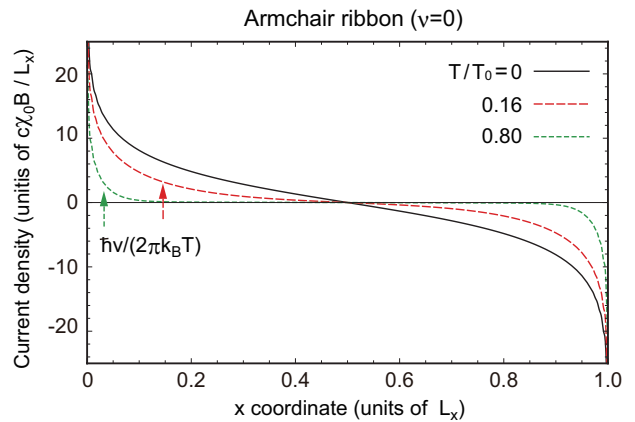}
\end{center}
\caption{
Diamagnetic current density $j_y(x)$
of metallic armchair ribbon $(\nu=0)$ with $\vare_F=0$, 
at several different temperatures. The vertical arrows indicate
the characteristic length scale $\hbar v/(2\pi k_B T)$
measured from $x=0$.
} \label{metalcurboke}
\end{figure}

\subsection{Diamagnetic current density}

Fig.\ \ref{allcur} shows the diamagnetic current density
$j_y(x)$ in different types of 
graphene ribbons at $\vare_F=0$ and $T=0$,
calculated in the first order perturbation of $B$.
The unit of current density is taken as $c \chi_0 B/L_x$.
The current flows in opposite directions
in the left-hand side and right-hand side of the ribbon, 
to make a magnetization perpendicular to the layer.
Reflecting the absence of the characteristic length scale,
the current distribution is not localized to the edge
but spread in the entire width in a form of
slowly-varying monotonic function.

In zigzag ribbons, the current density actually becomes absolute zero
at $x=0$ and $L_x$,
in accordance with the constraint argued in the previous section.
The current sharply drops to zero at the edges,
and some oscillatory feature remains around the edge 
due to a finite cut-off energy. 
When we increase the energy cut-off (not shown), 
the curve appears to slowly approach
a fixed curve having a discontinuous jump at the edges.
In the armchair ribbons, $j_y$ is not necessarily zero
but logarithmically diverges at the both edges.
The numerical calculation converges much more rapidly there,
since there is no discontinuity as in the zigzag case.

Fig.\ \ref{cur3d} is the two-dimensional density plot $j_y(x; \vare_F)$
of the diamagnetic current density of metallic armchair ribbon,
as a function of position $x$ (horizontal axis)
and  Fermi energy (vertical).
In increasing $\vare_F$, the current distribution 
begins to oscillate as a function of $x$, 
with a characteristic wave length of 
the order of $k_F = \vare_F/(\hbar v)$.

The temperature dependence of the current density at $\vare_F=0$
is shown in Fig.\ref{metalcurboke} for the same metallic armchair ribbon.
When $T$ becomes as large as $T_0$,
the current distribution is localized at the boundary
forming the counter edge currents.
This is the same temperature range where
the oscillation of $\chi$ disappears and the bulk limit is achieved.
The depth of the current distribution
is characterized by 
\begin{equation}
\lambda_\textrm{edge}(T) = \frac{\hbar v}{2\pi k_B T},
\label{eq_l_edge}
\end{equation}
which shrinks in increasing temperature.
With the band velocity of graphene, $v \approx 10^6$ m/s,
it is estimated as $\lambda_\textrm{edge} \approx [1/T(K)] \mu$m.

This behavior is intuitively explained using the plot of 
$j_y(x; \vare_F)$ in Fig.\ref{cur3d}.
The current density at a finite $T$ is obtained by 
integrating $j_y(x; \vare_F)$ in $\vare_F$
with the thermal averaging factor $-\partial f/\partial\vare$.
The current of the middle part of the ribbon
vanishes in averaging the oscillating function
in $\vare_F$, while the cancellation is not complete
only near the edges, since $j_y$ is always positive and negative
in left and right ends, respectively.
The similar temperature dependence of the 
current distribution is found in other types of ribbons
considered here. This suggests that, 
in any finite pieces of graphene with length scale $L$,
the finite-size effect disappears when $T > T_0$,
and then the diamagnetic current circulates only near edge
with a depth $\lambda_\textrm{edge}$.

\subsection{Relation to spin paramagnetism}

We neglect the effect of the electron spin through out the present analysis.
In a zigzag ribbon, particularly,
the large density of states contributed by the zero-energy
flat band is expected to give a significant magnitude of Pauli paramagnetism
and reduce 
the orbital diamagnetism. \cite{Wakabayashi_et_al_1999a}

The ratio between two effects can be quantitatively estimated as follows.
The susceptibility of Pauli paramagnetism is given by 
\begin{equation}
\chi_\textrm{para} = 
\left(\frac{g}{2}\right) \mu_B^2 D(\vare),
\end{equation}
where $g \sim 2$ is the $g$-factor for graphene electron,
$\mu_B = e\hbar/(2mc)$ is the Bohr-magneton
with $m$ being the free-electron mass, and
$D(\vare)$ is the density of states per area
given by the zero-energy flat band.
Since the number of edge 
states accommodated in a ribbon of the length $L$
is $\sim L/a$ per spin and per valley, \cite{Wakabayashi_et_al_1999a}
we have
\begin{equation}
D(\vare) \sim \frac{g_v g_s}{L^2} \frac{L}{a}\delta(\vare),
\end{equation}
which gives a delta-function singularity in $\chi_\textrm{para}$.

By comparing $\chi_\textrm{para}$ 
with the bulk orbital diamagnetism $\chi_\textrm{dia}$,
Eq.\ (\ref{eq_chi_zero_T}), we obtain
\begin{equation}
\Bigl| \frac{\chi_\textrm{para}}{\chi_\textrm{dia}}\Bigr|
= \frac{3\pi}{2}\frac{\hbar^2}{m^2 v^2 a L}
\sim 1.0 \times \frac{a}{L},
\end{equation}
which is negligible in a wide strip with $L \gg a$.
In a low temperature such that $k_B T <\sim \vare_0$, however,
$\chi_\textrm{dia}$ cannot be 
regarded as thermally broadened delta-function
due to the effect of the subband formation, and then 
$\chi_\textrm{para}$ 
overcomes $\chi_\textrm{dia}$ only at $\vare_F = 0$.

\section{Carbon nanotubes}
\label{sec_cnt}

The carbon nanotube is a quasi-one-dimensional
system similar to graphene ribbon,
but different in that there are no edges.
\cite{Iijima_1991a,Iijima_et_al_1992a}
Experimentally, graphene nanoribbons with smooth edges
can be obtained by unzipping the carbon nanotubes,
i.e., lengthwise cutting of carbon nanotube side walls. \cite{Kosynkin_et_al_2009a,Jiao_et_al_2009a}
Then we may ask which of the ribbon 
and the original nanotube has greater diamagnetism,
and how the susceptibility oscillation in $\vare_F$ changes in unzipping.
The orbital susceptibility of carbon nanotube
was theoretically studied for small Fermi energies 
in the effective mass approximation \cite{Ajiki_and_Ando_1993b,Ajiki_and_Ando_1995c}.
Here we compute full Fermi energy dependence 
in parallel fashion to the analysis for ribbons.

A carbon nanotube is characterized by a chiral vector,
\begin{align}
\Vec{L} = n_a \Vec{a} + n_b \Vec{b},
\end{align}
where the atom at $\Vec{L}$ on a graphene sheet
is rolled up onto the origin in constructing a tube.
The boundary condition is given by
$\psi_{\rm{A}}(\bm{R}_{\rm{A}})=
\psi_{\rm{A}}(\bm{R}_{\rm{A}}+\Vec{L})$ 
and 
$\psi_{\rm{B}}(\bm{R}_{\rm{B}})=
\psi_{\rm{B}}(\bm{R}_{\rm{B}}+\Vec{L})$.
For the effective mass wavefunction, it is written as \cite{Ando_2005a}
\begin{eqnarray}
\Vec{F}^\rm{K}(\Vec{r}+\Vec{L})= 
\exp\left(-\frac{2\pi i}{3}\nu\right)
\Vec{F}^\rm{K}(\Vec{r}),
\nonumber \\
\Vec{F}^\rm{K'}(\Vec{r}+\Vec{L})= 
\exp\left(+\frac{2\pi i}{3}\nu\right)
\Vec{F}^\rm{K'}(\Vec{r}).
\label{eq_bc_cnt}
\end{eqnarray}
Here $\Vec{F}^{\rm K} = (F_{\rm A}^{\rm K}, F_{\rm B}^{\rm K})$ etc.,
and $\nu$ is an integer $(0,\pm1)$ defined by
\begin{align}
 n_a + n_b = 3m + \nu,
\end{align}
with integer $m$.

For $K$ point, the eigenstates are immediately obtained as
\begin{align}
&\bm{F}^{\rm K}_{snk_y}(\bm{r})
=\frac{e^{{ik_yy}}}{\sqrt{2L_xL_y}}\begin{pmatrix}
e^{ik_n x}  \\
s e^{i(k_n x+\theta)} \\
0 \\ 0
\end{pmatrix}, \notag \\
&\ve_{snk_y}=s\hbar v\sqrt{k_y^2+k_n^2},
\end{align}
where 
\begin{align}
k_n \equiv 
\frac{2\pi}{L_x}
\left(n-\frac{\nu}{3}\right), {~}
n=0,\pm1,\pm2,\cdots,
\label{eq_kn_cnt}
\end{align}
and $L_x = |\Vec{L}|$ and $L_y$ is length of the carbon nanotube.
The system is metallic when $\nu=0$, and
semiconducting when $\nu=\pm 1$.
The band structure looks similar to armchair graphene ribbon's,
but the unit of momentum quantization doubled 
compared to Eq.\ (\ref{eq_kn_arm}),
leading to wider energy spacing between subbands.
The energy band for K$'$ is obtained by replacing $k_y$
by $-k_y$ and also $\nu$ by $-\nu$.

When a uniform magnetic field $B$ is applied 
perpendicularly to the nanotube axis, the vector potential
can be taken as
\begin{align}
\Vec{A}(\Vec{r}) = \left(0, \, 
\frac{B L_x}{2\pi}\sin \frac{2\pi x}{L_x}\right).
\end{align}
We should note that the expression differs from that for ribbon,
Eq.\ (\ref{eq_A}), because the magnetic field perpendicular to the 
tube surface is not a constant, but a sinusoidal function in $x$.
Except for that, the magnetic susceptibility $\chi_\textrm{tube}(\vare_F)$
is calculated in the same formula, Eq.\ (\ref{chi}).

The susceptibility of the carbon nanotube
is naturally related to that of graphene
against a spatial varying
magnetic field $B(q) \sin q x$ with $q=2\pi/L_x\equiv q_0$.
When we define the $q$-dependent susceptibility of graphene
as $\chi_\textrm{gr}(q) \equiv m(q)/B(q)$, \cite{Koshino_et_al_2009a}
we obtain a relation
\begin{align}
 \langle \chi_\textrm{tube}(\vare_F) \rangle_\varphi 
= \frac{1}{2}\chi_\textrm{gr}(q_0;\vare_F).
\label{eq_chi_cnt_av}
\end{align}
Here $ \langle \, \rangle_\varphi$ represents
an average over a phase factor $\varphi$ which twists the boundary condition
of carbon nanotube 
as $\psi(\Vec{r}+\Vec{L})=\exp(2\pi i \varphi) \psi(\Vec{r})$.
Physically, the phase factor 
corresponds to threading a magnetic flux of $(h/e)\varphi$ 
into the nanotube cross section.\cite{Ajiki_and_Ando_1993b,Ando_2005a}
It changes momentum quantization of Eq.\ (\ref{eq_kn_cnt})
to $k_n = (2\pi/L_x)(n+\varphi-\nu/3)$, 
and the flux averaging over $\varphi$ smears the difference in
$\nu$.
The factor $1/2$ in Eq.\ (\ref{eq_chi_cnt_av})
enters because  the average of
the squared  magnetic field on the nanotube surface is
$B(q_0)^2/2$.

At the zero temperature, $\chi_\textrm{gr}(q;\vare_F)$ 
is explicitly evaluated as \cite{Koshino_et_al_2009a}
\begin{eqnarray}
 \chi_\textrm{gr}(q;\vare_F)
&=& -\frac{g_vg_s e^2v}{16\hbar c^2} \, 
\frac{1}{q} \,  \theta(q-2k_F) \nonumber\\
&& \hspace{-20mm}\times
\left[
1
+ \frac{2}{\pi} \frac{2k_F}{q} \sqrt{1-\Big(\frac{2k_F}{q}\Big)^2}
- \frac{2}{\pi} \sin^{-1} \frac{2k_F}{q}
\right],
\label{eq_chi_gr}
\end{eqnarray}
where $k_F = |\vare_F|/(\hbar v)$ is the Fermi wave number
and $\theta(x)$ is 
defined by $\theta(x) = 1\,(x>0)$ and $0\,(x<0)$.
Using Eqs.\ (\ref{eq_chi_cnt_av}) and (\ref{eq_chi_gr}),
the flux-averaged susceptibility integrate is shown to be
\begin{align}
 \Bigl\langle 
\int_{-\infty}^\infty \chi_\textrm{tube}(\vare_F) d\vare_F 
\Bigr\rangle_\varphi 
= \frac{1}{2}\left(
-g_vg_s \frac{e^2 v^2}{6\pi c^2} 
\right),
\end{align}
which is exactly half of graphene's,
suggesting that the susceptibility is effectively smaller in nanotube
than in ribbon.
This is simply because the $B$-field component penetrating the lattice plane
is smaller in the nanotube 
due to its cylindrical shape.

\begin{figure}[b]
\begin{center}
\leavevmode\includegraphics[width=0.9\hsize]{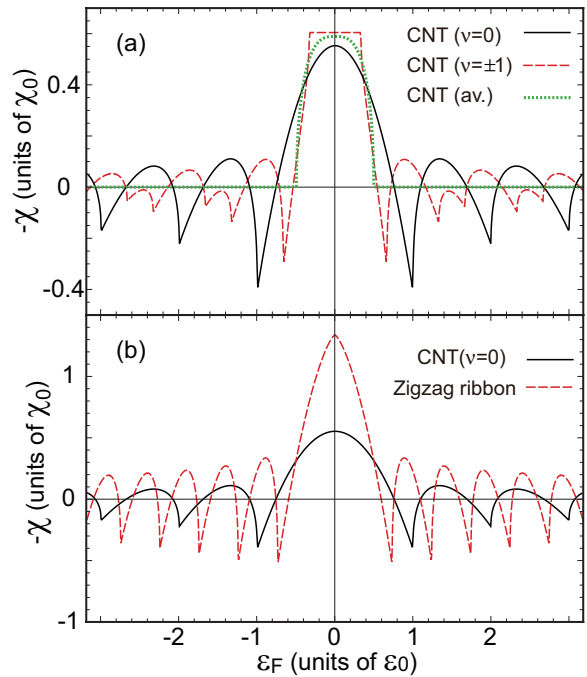}
\end{center}
\caption{(a) Magnetic susceptibility of 
carbon nanotubes. Solid (black), dashed (red)
and dotted (green) curves are
for metallic ($\nu=0$),
semiconducting ($\nu=\pm 1$),
and the flux average, respectively.
(b) Magnetic susceptibility of 
the carbon nanotube of $\nu=0$ (solid black)
and the zigzag ribbon unzipped from the same nanotube
(dashed red).
} \label{tube}
\end{figure}

The susceptibility before taking flux average
can be calculated in numerics.
Fig.\ \ref{tube} (a) shows $\chi_\textrm{tube}(\vare_F)$
for the metallic ($\nu=0$)
and the semiconducting ($\nu=\pm 1$) nanotubes,
together with the flux average.
It has an oscillatory behavior similar to the graphene ribbon's,
while $\chi$ in $|\vare_F| > \hbar v q /2$
completely vanishes after flux average.\cite{Koshino_et_al_2009a}
In increasing temperature (not shown), 
the oscillation immediately disappears,
leaving a single peak regardless of $\nu$, similar to the graphene ribbon.

Fig.\ \ref{tube} (a) compares the susceptibility of 
a carbon nanotube and that of corresponding graphene ribbon
unzipped from the same nanotube.
Here we chose a zigzag ribbon as an example,
when the corresponding nanotube becomes
an armchair nanotube which is always metallic $(\nu=0)$. \cite{Ando_2005a}
The oscillation period of the nanotube
is approximately twice as large as that of the ribbon,
reflecting the wider subband spacing.
Overall magnitude of $\chi$ is smaller
in nanotube roughly by factor 2.
The integrate of susceptibility in $\vare_F$
differs in factor 2 in numerical accuracy,
in accordance with the above arguments.

\section{Randomly-stacked multilayer graphene}
\label{sec_random}

The diamagnetism can be made
even greater by stacking graphenes in three dimensions.
The recent experimental technique
realizes a novel kind of graphene multilayer
in which successive layers are stacked
with random rotating angles. \cite{Berger_et_al_2006,Hass_et_al_2007a,G.Li_et_al_2009a}
There it is known that the interlayer coupling is significantly weakened
and the Dirac cone is kept almost intact near zero energy
as long as the rotating angle is not too small. 
\cite{Lopes_dos_Santos_et_al_2007a,Latil_et_al_2007a,Shallcross_et_al_2008a,Mele_2010a}
The orbital susceptibility of such a system is expected to 
be much stronger than graphite in which 
the delta-function peak of $\chi(\vare_F)$ 
is much broadened and shortened by the regular interlayer coupling. 
\cite{Safran_1984a,Koshino_and_Ando_2007c}

Here we consider the orbital diamagnetism of 
a finite-sized piece of random-stacked graphene multilayers.
In calculations, 
we self-consistently include the effect of the counter magnetic field 
induced by the diamagnetic current itself.
This is essential because, as we will show in the following,
the counter magnetic field of this system
can be of the same order of the external magnetic
field, and even nearly perfect screening is possible in low temperatures.

For simplicity, we completely neglect the interlayer coupling
and regard the system as a set of independent single layer graphenes.
We also assume that each layer has the identical shape
with a characteristic length scale $L$,
and that the system is large enough that
the thermal broadening energy $k_B T$ 
is much larger than $2\pi\hbar v/L$.
According to the previous discussions,
we then expect that the susceptibility of each layer 
is given by the bulk limit $\chi_\textrm{gr}$
in Eq.\ (\ref{eq_chi_in_T}),
and also the depth of the edge current $\lambda_\textrm{edge}$
of Eq.\ (\ref{eq_l_edge})
can be neglected with respect to the system size $L$.

Let us consider a situation where
a external field $B_\textrm{ext}$ is applied perpendicularly to
graphene plane of the random stacked multilayer.
The total magnetic field $B$ penetrating the system is
\begin{align}
 B = B_\textrm{ext} + \Delta B,
\end{align}
where $\Delta B$ is the counter field caused by
graphene electrons. The total field $B$ induces the magnetism
in each layer,  
$M = \chi_\textrm{gr} B S$, with $S$ being the area of the layer.
This is related to the diamagnetic edge current $I$
of each single layer by
\begin{align}
 I  = \frac{c M}{S} = c \chi_\textrm{gr} B.
\end{align}
Since the ring current $I$ exists every interlayer distance $d$,
it induces a counter-magnetic field inside the system as
\begin{align}
 \Delta B = \frac{4\pi}{c} \frac{I}{d}.
\end{align}

Solving the set of equations, we find that 
the dimensionless volume susceptibility becomes
\begin{align}
 \chi_\textrm{3D} \equiv \frac{\Delta B}{B_\textrm{ext}}
= \frac{-1}{1 - d/(4\pi\chi_\textrm{gr})}.
\end{align}
At the charge neutral point $\mu=0$, in particular, 
we have
 \begin{align}
 \chi_\textrm{3D}(\mu=0) 
= \frac{-1}{1 + k_B T/\Delta},
\end{align}
where $\Delta$ is a characteristic energy scale defined by
\begin{equation}
\Delta = \frac{g_vg_s}{6}\left(\frac{v}{c}\right)^2\frac{e^2}{d}
\approx 0.03 \textrm{meV},
\end{equation}
and $d$ is assumed to be the interlayer spacing of graphite,
0.334 nm.

In decreasing the temperature, 
$\chi_\textrm{3D}$ monotonically increases in the negative direction,
and approaches $-1$, where the prefect magnetic field screening is achieved.
This reflects the property of the
single-layer susceptibility, Eq.\ (\ref{eq_chi_in_T}), 
of which peak value at Dirac point diverges
in $T \to 0$.
In contrast, $\chi_\textrm{3D}$ of the graphite is of the order of
$10^{-4}$ and is not much 
enhanced in low temperatures, \cite{Krishnan_Ganguli_1937a} 
because $\chi(\vare_F)$ is already
broadened by the interlayer coupling energy
about $4000$ K. \cite{Safran_1984a,Koshino_and_Ando_2007c}
A three-dimensional bulk material composed of random-stacked
graphenes, if realized, would be the strongest diamagnetic material
than any other known substances except for the superconductors.
Including the effect of the residual 
interlayer coupling between misoriented layers
may set the upper limit to $\chi_\textrm{3D}$,
while we leave the detailed analysis for a future problem.

\section{Conclusion}
\label{sec_concl}

We have studied the orbital diamagnetism of
graphene ribbons using the effective-mass approximation, 
to figure out its dependence on temperature and Fermi energy,
and also the finite-size effect on the delta-function singularity 
in the bulk limit.
In increasing temperature, an oscillatory behavior in
the orbital susceptibility $\chi(\vare_F)$
is eventually smeared out, approaching the bulk limit
i.e., a thermally broadened delta-function centered at $\vare_F=0$.
The electric current responsible for the diamagnetism 
spreads entirely in the sample at $T=0$
reflecting the absence of the characteristic wavelength,
while as $T$ is increased, the current density tends to localize 
near the edge with a depth  $\sim \hbar v /(2\pi k_B T)$.

We also see a carbon nanotube, another form
of quasi-one-dimensional carbon, exhibits a similar 
oscillation in $\chi(\vare_F)$, but the overall magnitude
is reduced by a factor 2 compared to the corresponding 
ribbon having the same width.
The result is applied to estimate the three-dimensional bulk 
susceptibility of random-stacked multilayer graphene. 
There we showed that the external magnetic field is significantly screened 
inside the sample.

\section*{ACKNOWLEDGMENTS}
This project has been funded by JST-EPSRC Japan-UK
Cooperative Programme Grant No. EP/H025804/1.

\end{document}